\documentclass[conference]{IEEEtran}

\usepackage{cite}
\usepackage{amsmath,amssymb,amsfonts}
\usepackage{algorithmic}
\usepackage{graphicx}
\usepackage{textcomp}
\usepackage{xcolor}
\usepackage{algorithm}
\usepackage{algorithmic}
\usepackage{subfigure}
\newtheorem{definition}{Definition}
\newtheorem{proposition}{Proposition}

\makeatletter
\newcommand{\removelatexerror}{\let\@latex@error\@gobble}
\makeatother

\def\BibTeX{{\rm B\kern-.05em{\sc i\kern-.025em b}\kern-.08em
    T\kern-.1667em\lower.7ex\hbox{E}\kern-.125emX}}

\begin{document}
    \title{QoS-Driven Video Uplinking in NOMA-Based IoT}

    \author{\IEEEauthorblockN{Pengfei Ma, Hancheng Lu, Ming Zhang, Jinxue Liu, Ruoyun Chen}
    \IEEEauthorblockA{
        The Information Network Lab of EEIS Department, USTC, Hefei, China, 230027 \\
        Email: mpf916@mail.ustc.edu.cn, hclu@ustc.edu.cn, mzhang95@mail.ustc.edu.cn,\\ jxliu18@mail.ustc.edu.cn, chenryun@mail.ustc.edu.cn
    }}
    
    \maketitle

    \begin{abstract}
        In recent years, with the explosive growth of visual sensors and a large number of related video applications in Internet of Things (IoT), massive video data is generated by IoT devices. 
        Since the volume of video data is far greater than traditional data in IoT, it is challenging to ensure high Quality of Service (QoS) for video uplinking in IoT. 
        To address this challenge, we integrate non-orthogonal multiple access (NOMA) and scalable video coding (SVC) in IoT. 
        To improve the video quality, we formulate a power allocation problem to maximize the average QoS in the proposed integrated system. 
        Due to that the problem is non-convex, we transform it into a monotonic problem based on its hidden monotonicity. 
        Then a power allocation algorithm based on polyblock outer approximation is proposed to solve the problem effectively. 
        Finally, simulation results demonstrate that the proposed algorithm outperforms existing OMA and NOMA based schemes for video uplinking in IoT in terms of QoS and energy efficiency. 
    \end{abstract}


    \section{Introduction}
        Internet of Things (IoT) provides a seamless connection between the cyberspace and the real world. 
        With the rapid development of IoT applications in recent years, the IoT devices have transformed from simple sensors to more and more complicated devices. Among them, multimedia devices with visual sensors, such as audio-visual cameras, high definition video cameras and camera arrays, play an important role in current IoT. 
        As a result, the vigorous growth of the multimedia devices and massive multimedia contents generated by the devices has given rise to the emergence of several new concepts in IoT, such as Internet of Video Things (IoVT), Visual IoT, Multimedia IoT, which need a reliable transmission scheme for video uplinking\cite{chen2020internet,ji2019visual,nauman2020multimedia,fu2020secure}. 
        Video uplinking can be used in plenty of applications, including real-time traffic/facility monitoring, damage assessment, disaster relief and so on. 
        However, the limited bandwidth and energy make massive video data uplinking face huge challenges. 

        In order to achieve high-quality video transmission, both spectrum efficiency (SE) and energy efficiency (EE) of the wireless network should be further improved\cite{chen2020internet}. 
        However, traditional orthogonal multiple access (OMA) can not effectively support the video transmission of a large number of visual sensors. 
        In recent few years, non-orthogonal multiple access (NOMA) has been regarded as the promising technology to greatly improve SE\cite{dai2015non,gong2014virtual}. 
        Different from the OMA technology, NOMA provides access to multiple users in the power domain through superposition coding (SC) and successive interference cancellation (SIC) technologies.
        In uplink NOMA systems, different users transmit information at the same time-frequency resource block through SC, while the base station (BS) decodes the signal from different users with SIC. 
        Therefore, NOMA can be used to get better SE performance, which is a promising approach for video uplinking in IoT. 
        To make NOMA serve the video uplinking better, there are still some challenges to solve. 
        First, to serve the applications better i.e., market/facility monitoring, disaster relief \textit{et al.}, a proper video quality model is required to ensure that we can optimize the QoS effectively. 
        Second, due to the interference among different devices, how to allocate the transmission power for these devices needs to be considered carefully. 
        Specially, we need to design an effective power allocation algorithm to make full use of the limited transmission energy with the assistance of the video quality model. 

        Due to the advantage of NOMA, it has attracted great research interests with the extensive application to IoT and video transmission. 
        Considering different kind of IoT devices, a NOMA based NB-IoT system is proposed to improve the connectivity of IoT devices\cite{mostafa2019connection}. 
        The authors in \cite{wang2020throughput} propose a peer-assisted power supply approach for a wireless powered IoT NOMA network. 
        In \cite{shahini2019noma}, the authors propose a power domain NOMA scheme with user clustering in an NB-IoT system with MTC devices. 
        However, these studies only focus on the connectivity and throughput for data transmission, which can not satisfy the QoS requirement of video uplinking in IoT. 
        When it comes to video transmission in NOMA systems, a multi-user video unicast scheme is proposed for NOMA system in \cite{jiang2018enabling}, while Zhang \textit{et al.} present a scalable video multicast scheme to improve the quality of experience (QoE) of all broadcast users with NOMA\cite{zhang2020noma}. 
        In \cite{lu2020qoe-driven}, the authors propose a spatial modulation (SM) and NOMA integrated system for multi-user video transmission. 
        A cross-layer scalable video delivery scheme is proposed in \cite{lu2020distortion}, which takes distortion and the content characteristics at the application layer into consideration. 
        There are also studies focusing on improving the quality of video in NOMA based networks\cite{zhu2018scalable,xu2018max,li2019noma,lu2015highly}. 
        Even so, these approaches are not applicable for video uplinking in NOMA systems. 
        To improve the multimedia upstreaming service in the ever-growing live-streaming IoT market, a game theoretical power allocation solution is proposed in \cite{he2019multimedia} for multimedia IoTs with modern NOMA capabilities. 

        However, the model in it doesn't take the characteristics of video into consideration so that it fails to improve the quality of the video as expected and to be energy efficiency at the same time. 
        Another flaw of the existing work is that none of these approaches takes the energy efficiency constraints into consideration, which is important in IoT\cite{chen2020internet}. 
        So far, all the aforementioned works can not solve the challenges of video uplinking in NOMA-baesd IoT well. 

        To address the challenges above, we introduce NOMA for video uplinking in IoT, which can provide high SE to alleviate the problem of limited spectrum. 
        We adopt a discrete QoS model based on SVC to provide the best video with flexible and
        variable quality under different channel conditions, which make the scheme more suitable for video transmission, 
        The main contributions of this paper are summarized as follows: 
        \begin{itemize}
            \item We propose to integrate NOMA and SVC and formulate a QoS-driven power allocation problem to provide optimal QoS services. 
            \item To tackle this non-convex optimization problem, we prove its hidden monotonicity and transform it into a monotonic problem. Then we propose an optimal power allocation algorithm based on polyblock outer approximation. 
            \item Extensive simulations are carried out to validate our scheme. The performance of the proposed algorithm is evaluated under different power constraints and different distribution ranges of devices. The results reveal that the proposed algorithm shows a better performance than existing NOMA and OMA based schemes. 
        \end{itemize}

        The rest of the paper is organized as follows.
        The system model is presented in Sec. \uppercase\expandafter{\romannumeral2}. 
        The problem analysis and formulation are given in Sec. \uppercase\expandafter{\romannumeral3}. 
        In Sec. \uppercase\expandafter{\romannumeral 4}, we design an optimal power allocation algorithm based on monotonic optimization. 
        Performance of our proposed scheme is evaluated in Sec. \uppercase\expandafter{\romannumeral 5}. 
        Sec. \uppercase\expandafter{\romannumeral 6} concludes our work in this paper. 

    \section{System Model}

\begin{figure}[hbp]
            \centerline{\includegraphics[width = .7\columnwidth]{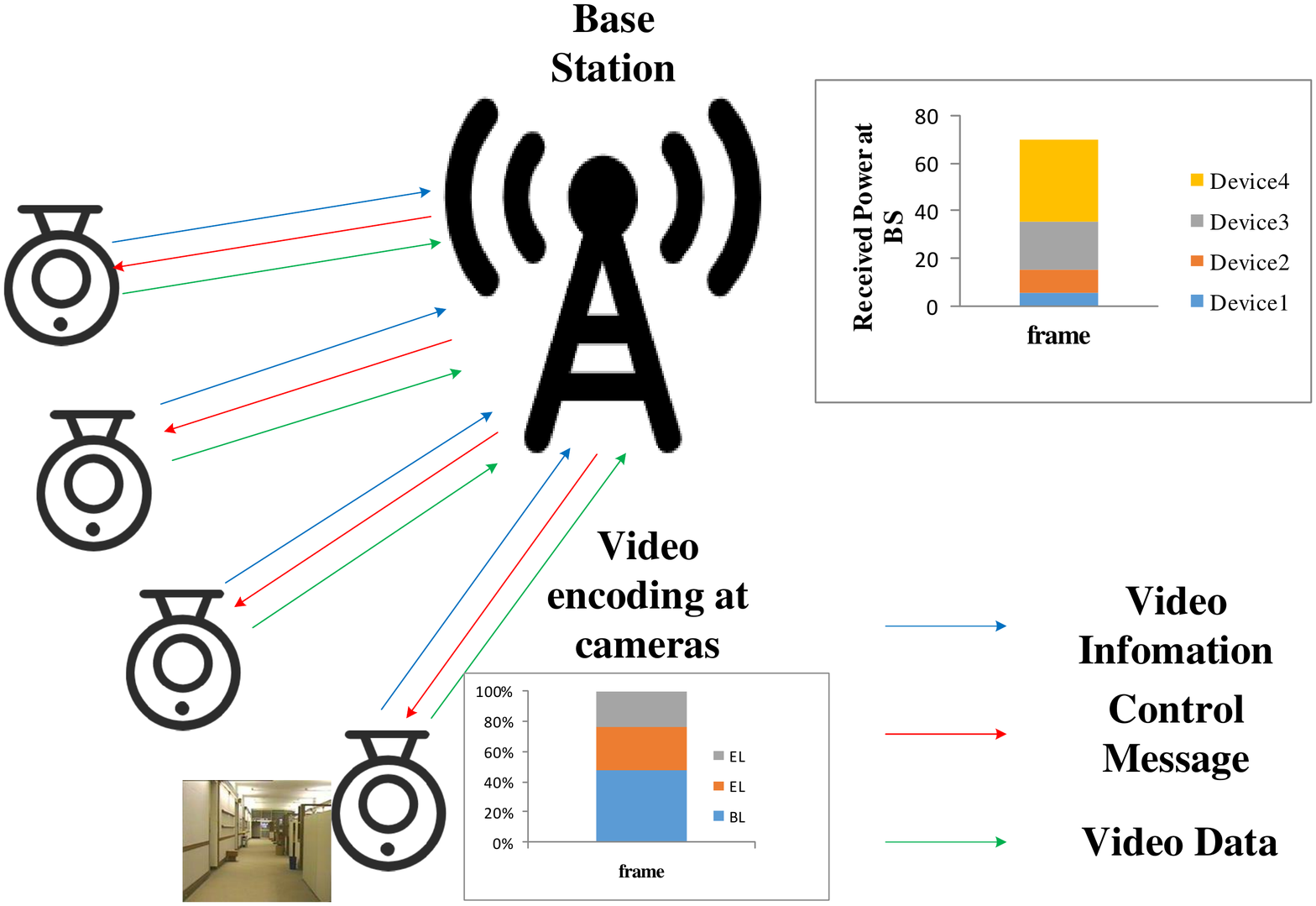}}
            \caption{Architecture of Video Uplinking in NOMA-Based IoT}
            \label{fig:sys-mod}
        \end{figure}

        Video uplinking in NOMA-based IoT is considered in this paper. 
        In this model, there are one base station (BS) and several IoT devices (wireless visual sensors) whose architecture is shown in Fig.~\ref{fig:sys-mod}. 
        During the video uplinking, several devices upload their captured video to the BS simultaneously using the same spectrum resource through SC. 
        As the receiver, the BS decodes the video signals of different devices through SIC. 
        In order to reduce power consumption as much as possible while improving the QoS of the uploaded video, we propose a video upload scheme that reports information before transmitting the video. Before the transmission, each device sends the video information (the bit rate required for different video quality) and its power constraints to the BS. 
        Then BS calculates a power allocation plan based on the obtained information and sends it back to the devices. 
        Subsequently, the devices simultaneously send video data to the BS according to the allocated transmit power and the corresponding achievable bit rate. 
        
    \subsection{NOMA-based Video Transmission Model}

        In the video uplinking system of NOMA-based IoT, we suppose that there are a total of $M$ IoT devices (visual sensors), expressed as: $D=\{D_i \vert i=1,2,\cdots,M\}$. Without loss of generality, we assume that they have been arranged in descending order of the channel gain $h_i$. 
        The channel gain between the BS and device $D_i$ is $h_i$, the video signal sent by the device $D_i$ is $s_i$, and the transmitting power is $p_i$. Then, the signal received by BS is 
        \begin{equation}
            \label{eq:rcv-sig}
            y=\sum\limits_{i=1}^M h_i \sqrt{p_i}s_i+\omega,
        \end{equation}
        where $\omega$ is additive white Gaussian noise with zero mean and variance $\sigma_\omega^2$. 
        At the BS, the signal can be decoded by SIC in the descending order of channel gain. 
        Starting from the signal of $D_1$, the signal of $D_i$ is decoded using SIC technology, and then the decoded signal is eliminated from the remaining signal to decode $D_{i+1}$'s signal. 

        According to this decoding order, when decoding the signal of the device $D_i$, the signal of the device with the index greater than $i$ is regarded as the interference. 
        Therefore, the signal to interference and noise ratio (SINR) of device $D_i$ is 
        \begin{equation}
            \label{eq:sinr}
            \gamma_i=\frac{h_i^2 p_i}{\sum_{j=i+1}^M h_j^2 p_j +\sigma_\omega^2}.
        \end{equation}

        According to the Shannon equation, the achievable transmission rate of $D_i$ is 
        \begin{equation}
            \label{eq:rate}
            R_i=B \log_2 (1+\gamma_i) =B \log_2 (1+\frac{h_i^2 p_i}{\sum_{j=i+1}^M h_j^2 p_j +\sigma_\omega^2}),
        \end{equation}
        where $B$ denotes the channel bandwidth in unit of Hz. 

        After obtaining the rate and transmit power, the energy efficiency of $D_i$ can be calculated by 
        \begin{equation}
            \label{eq:ee}
            e_i=\frac{R_i}{p_i}.
        \end{equation}

    \subsection{QoS Model}
        In order to provide video data with different quality and different bit rate, we use SVC to encode the video into multiple layers. 
        In this way, the wireless visual sensors can send videos of different quality according to the achievable transmission rate calculated by Eq.~\eqref{eq:rate}. 

        For the video captured by device $D_i$, it can be encoded into $L_i$ layers $\{s_{i,l}\}_{l=1}^{L_i}$ by SVC, which contains a base layer $s_{i,1}$ and $L_i-1$ enhancement layers, $s_{i,l}$ means the $l$-th layer of the video. 
        Assuming that $R_{i,l}$ represents the bit rate required by $D_i$ to send the first layer to the $l$-th layer of the video, the quality of the video reconstructed with these $l$ layers compared with the original video can be measured by the peak signal-to-noise ratio (PSNR), which depends on the mean squared error (MSE) given by 
        \begin{equation}
            \label{eq:mse}
            MSE_{i,l}=\frac{1}{W_i H_i} \sum_{x=0}^{W_i-1}\sum_{y=0}^{H_i-1} \vert f(x,y)-g(x,y) \vert ^2,
        \end{equation}
        where $W_i$ and $H_i$ are the width and height of the frames, $f(x.y)$ is the pixel's luminance intensity in the original frame and $g(x,y)$ is the pixel's luminance intensity in the reconstructed frame. 

        And the PSNR is defined as follows 
        \begin{equation}
            \label{eq:Psnr}
            Q_{i,l}=PSNR_{i,l}=10 \log_{10} \frac{255^2}{MSE_{i,l}}.
        \end{equation}
        $Q_{i,l}$ is the PSNR when the device sends $l$ layers of the video. 
        In this way, we can collect the relationship between rate and the quality of received video as a set $\{(R_{i,l},Q_{i,l})\}$. 
        For a certain $R_i$, we can transmit the video of the corresponding quality (PSNR) with the $R_{i,l}$ which is not higher than $R_i$. 

    \section{Problem Formulation}

        Based on the QoS model in Sec. \uppercase\expandafter{\romannumeral 2}, we can calculate the contribution of each video layer to the video's QoS. 
        As $s_{i,l}$ represents the $l$-th layer of the video, the contribution of $s_{i,l}$ is defined as the difference between the PSNR of the video that contains $s_{i,l}$ and one that contains $s_{i,l-1}$, which can be expressed as 
        \begin{equation}
            \label{eq:psnr}
            q_{i,l}=Q_{i,l}-Q_{i,l-1}.
        \end{equation} 

        With the definition in Eq.~\eqref{eq:psnr}, we can obtain the PSNR function of power $\boldsymbol{p}$ 
        \begin{equation}
            \label{eq:PSNR}
            Q_i(\boldsymbol{p})=\sum_{l=1}^{L_i} q_{i,l}I(R_i(\boldsymbol{p}) \geq R_{i,l}),
        \end{equation}
        where $I(\cdot)$ is an indicator function that returns 1 if the argument is true and 0 otherwise. And $R_i$ can be calculated with Eq.~\eqref{eq:rate} and the power vector $\boldsymbol{p}=(p_1,p_2,\cdots,p_M)$. 

        To make sure that the BS can receive the video of every visual sensor, it should be ensured that the BS can receive the video of the lowest quality, which is the base layer of the video. 
        To achieve that, we guarantee that the achieved rate is higher than or equal to the rate required by the base layer of the video. 
        It is expressed as 
        \begin{equation}
            \label{eq:cons-rate}
            R_i \geq R_{i,1},\forall i \in \mathcal{M}.
        \end{equation}

        As for devices in IoT, they should be energy efficiency constrained. 
        Based on the definition in Eq.~\eqref{eq:ee}, we formulate the energy efficiency constraints as 
        \begin{equation}
            e_i \geq e_i^{min},\forall i \in \mathcal{M}.
        \end{equation}

        In summary, we formulate the power allocation for each wireless visual sensors to maximize the average QoS of the system while guaranteeing the basic video transmission for all devices and also, taking the energy efficiency constraints into consideration. 
        The final power allocation problem is formulated as follows: 
        \begin{align}
            \max\limits_{\boldsymbol{p}} \quad &\frac{1}{M}\sum\limits_{i=1}^M Q_i(\boldsymbol{p}), \label{prob-1}\\
            s.t.\quad & R_i \geq R_{i,1},\forall i \in \mathcal{M} \tag{\ref{prob-1}a},\label{cons-1a}\\
            \quad & 0 \leq p_i \leq p_i^{max},\forall i \in \mathcal{M} \tag{\ref{prob-1}b},\label{cons-1b}\\
            \quad & e_i \geq e_i^{min},\forall i \in \mathcal{M} \tag{\ref{prob-1}c},\label{cons-1c}
        \end{align}
        where $\mathcal{M}=\{1,2,\cdots,M\}$ and $\boldsymbol{p}=\{p_i \vert i \in \mathcal{M}\}$. 
        \eqref{cons-1b} is a non-negative transmission power constraint with a maximum transmission power. 

        Due to the non-convexity and discontinuity of the objective function, it is not easy to find a global optimal solution for problem \eqref{prob-1}. 
        But we can utilize the hidden monotonicity of problem \eqref{prob-1} to design an optimal power allocation algorithm, which will be discussed in the next section. 

    \section{Solution to the optimization problem}
        In this section, we first introduce some basic definitions in monotonic optimization. 
        Then we prove that problem \eqref{prob-1} is a monotonic optimization problem. 
        Finally, we propose an optimal power allocation algorithm to maximize the QoS. 

        Some necessary definitions are introduced as follows\cite{zhang2013monotonic}: 
        \begin{definition}
            (Box) Given any vector $\boldsymbol{a},\boldsymbol{b} \in \mathcal{R}_+^N$, a box $[\boldsymbol{a},\boldsymbol{b}]$ is a set of $\{\boldsymbol{z}|\boldsymbol{a} \preceq \boldsymbol{z} \preceq \boldsymbol{b}\}$ if $\boldsymbol{a} \preceq \boldsymbol{b}$. 
            A box is also referred to a hyper-rectangle. 
        \end{definition}

        \begin{definition}
            (Normal set and Conormal set): A set $\mathcal{G} \subset \mathcal{R}_+^N$ is normal if for any vector $\boldsymbol{x} \in \mathcal{G}$, the box $[\boldsymbol{0},\boldsymbol{x}] \subset \mathcal{G}$. 
            A set $\mathcal{H}$ is conormal if $\boldsymbol{x} \in \mathcal{H}$ and $\boldsymbol{x}'\succeq \boldsymbol{x}$ implies $\boldsymbol{x}' \subset \mathcal{H}$. 
        \end{definition}

        The constant \eqref{cons-1a} can be transformed into SINR constraint $\gamma_i \geq \gamma_i^{min},i \in \mathcal{M}$, and we can reformulate the problem \eqref{prob-1} as $Q(\boldsymbol{\gamma}(\boldsymbol{p}))=\frac{1}{M}\sum_{i=1}^M Q_i(\boldsymbol{p})$. 
        The optimization problem can be rewritten as: 
        \begin{align}
            \max\limits_{\boldsymbol{p}} \quad &Q(\boldsymbol{\gamma}(\boldsymbol{p})), \label{prob-2}\\
            s.t.\quad&  \gamma_i \geq \gamma_i^{min},i \in \mathcal{M} \tag{\ref{prob-2}a},\label{cons-2a}\\
                \quad & 0 \leq p_i \leq p_i^{max},\forall i \in \mathcal{M} \tag{\ref{prob-2}b},\label{cons-2b}\\
                \quad & e_i \geq e_i^{min},\forall i \in \mathcal{M} \tag{\ref{prob-2}c}.\label{cons-2c}
        \end{align}

        The objective function is a monotonically increasing function of SINR $\boldsymbol{\gamma}(\boldsymbol{p})$. 
        We can obtain the optimal power allocation once we find the optimal $\boldsymbol{\gamma}$ according to Eq.~\eqref{eq:sinr}. 
        Now, we can rewrite the problem in the standard form of monotonic optimization problem: 
        \begin{equation}
            \begin{aligned}
                \max\limits_{\boldsymbol{y}} &\quad Q(\boldsymbol{y}) \label{prob-3}\\
                s.t. &\quad \boldsymbol{y} \in \mathcal{G} \cap \mathcal{H}
            \end{aligned}
        \end{equation}
        where the feasible set $\mathcal{G} \cap \mathcal{H}$ is obtained from constants \eqref{cons-2a}-\eqref{cons-2c}: 
        \begin{equation}
            \begin{aligned}
                \mathcal{G}&=\mathcal{G}_1 \cap \mathcal{G}_2\\
                \mathcal{G}_1&=\{\boldsymbol{y} \vert 0 \leq y_i \leq \gamma_i(\boldsymbol{p}),\forall i \in \mathcal{M},\boldsymbol{0} \preceq \boldsymbol{p} \preceq \boldsymbol{p^{max}}\}\\
                \mathcal{G}_2&=\{\boldsymbol{y} \vert 0 \leq y_i \leq \gamma_i(\boldsymbol{p}),\forall i \in \mathcal{M},\boldsymbol{e} \succeq \boldsymbol{e^{min}}\}\\
                \mathcal{H}&=\{\boldsymbol{y} \vert y_i \geq \gamma_i^{min},\forall i \in \mathcal{M}\}
            \end{aligned}
        \end{equation}
        According to \cite{zhang2013monotonic}, the set $\mathcal{G}_1$ is normal and the set $\mathcal{H}$ is conormal. 
        Now we prove that $\mathcal{G}_2$ is normal. 
        \begin{proposition}
            \label{prop-1}
            A set $\mathcal{F} \subset \mathcal{R}_+^N$ whose boundary is the surface $0=f(x_1,x_2,\cdots,x_N)$ is normal if $\frac{\partial x_1}{\partial x_i}\leq 0, \forall 2 \leq i \leq N$. 
        \end{proposition}
        \begin{IEEEproof}
            For the boundary of set $\mathcal{F}$, $\frac{\partial x_1}{\partial x_i}\leq 0, \forall 2 \leq i \leq N$, then for any $i,j>1$ at point $\boldsymbol{x}$ on the surface, the normal vector of the tangent to the parallel coordinate plane $x_iOx_j$ is $[\frac{\partial x_1}{\partial x_i},\frac{\partial x_1}{\partial x_j}]$. 
            As $\frac{\partial x_1}{\partial x_i},\frac{\partial x_1}{\partial x_j} \leq 0$, the slope of the tangent is $\frac{\partial x_i}{\partial x_j} \leq 0$. 
            Therefore, for any $1 \leq i,j \leq N$, $\frac{\partial x_i}{\partial x_j} \leq 0$. 
            We can also rewrite the equation of the surface as $x_i=g_i(x_1,\cdots,x_{i-1},x_{i+1}\cdots,x_N)$. 

            Based on that, $\forall \boldsymbol{x} \in \mathcal{F}$, assuming that $\boldsymbol{x'} \in [\boldsymbol{0},\boldsymbol{x}]$. 
            Since $\boldsymbol{x}' \preceq \boldsymbol{x}$, so $x'_i \leq x_i \leq g_i(x_1,\cdots,x_{i-1},x_{i+1}\cdots,x_N) \leq g_i(x'_1,\cdots,x'_{i-1},x'_{i+1}\cdots,x'_N)$.
            In conclusion, set $\mathcal{F}$ is normal because $\boldsymbol{x}' \in \mathcal{F}$.
        \end{IEEEproof}

        Then we will prove that the set $\mathcal{G}_2$ is a constant set (it will be proved by mathematical induction).
        \begin{IEEEproof}
            First consider only the energy efficiency constraint $e_M \geq e_M^{th}$ of the last decoded device $D_M$, then for the first $M-1$ devices, assuming that the feasible range of$\gamma_i(\boldsymbol{p}),1 \leq i \leq M-1$ is $R_+$ for the first $M-1$ device. For $M$-th device, there are $\frac{\log_2 (1+\frac{h_M^2 p_M}{\sigma_\omega^2})}{p_M} \geq e_M^{min}$. 
            From this inequality, we know that $\exists P_M > 0,0 \leq p_M \leq P_M$. According to the formula of SINR, $\exists \Gamma_M > 0,0 \leq p_M \leq \Gamma_M$. 
            So, the feasible region of SINR is $\mathcal{F}_M=\{\boldsymbol{\gamma}|\gamma_i \in \mathcal{R}_+,1 \leq i \leq M-1,0 \leq p_M \leq \Gamma_M\}$, obviously the set $\mathcal{F}_M$ is a normal. 

            Assuming that the SINR feasible region $\mathcal{F}_{i+1}$ is still normal when considering the energy efficiency constraints from $\{D_j\}_{j=i+1}^M$, 
            then the feasible region of SINR after introducing the energy efficiency constraints of the device $D_i$ is $\mathcal{F}_i=\mathcal{F}_{i+1} \cup \mathcal{F}'_i$ (where $\mathcal{F}'_i$ is the feasible region when the energy efficiency constraints of the device $D_i$ are considered separately). 
            The boundary of the set $\mathcal{F}'_i$ is a curved surface determined by energy efficiency constraints. 
            We cannot directly obtain the boundary surface equation of the SINR domain from the energy efficiency constraint, but we can obtain the surface equation of the power domain through the energy efficiency constraint and thus obtain some properties of the surface of the SINR domain. 
            The surface equation obtained from the energy efficiency constraint is 
            \begin{equation}
                \frac{\log_2(1+\frac{h_i^2 p_i}{\sum_{j=i+1}^M h_j^2 p_j +\sigma_\omega^2})}{p_i} = e_i^{min}.
            \end{equation}
            Derivation of this surface can be obtained, $\frac{\partial p_i}{\partial p_j} \leq 0,i+1 \leq j \leq M$. 
            According to the formula $\gamma_i=\frac{h_i^2 p_i}{\sum_{j=i+1}^M h_j^2 p_j +\sigma_\omega^2}$, $\frac{\partial \gamma_i}{\partial p_i}>0$ can be obtained. 
            From the previous derivative relationship and the SINR formula, it can be deduced that $\frac{\partial \gamma_i}{\partial p_j}<0,i+1 \leq j \leq M$, so $\frac{\partial \gamma_i}{\partial \gamma_j}=\frac{\partial \gamma_i}{\partial p_j}\frac{\partial p_j}{\partial \gamma_j}<0,i+1 \leq j \leq M$. 
            According to proposition (\ref{prop-1}), the set $\mathcal{F}'_i$ is normal. 
            So set $\mathcal{F}_i$ is still normal. 
            Therefore, set $\mathcal{G}_2$ is normal.
        \end{IEEEproof}

        Now we can conclude that set $\mathcal{G}$ is normal, and problem \eqref{prob-3} is a standard form of monotonic optimization problem. 
        We design a power allocation algorithm based on polyblock outer approximation (POA) approach\cite{zhang2013monotonic}. 
        A polyblock is a union of boxes $[\boldsymbol{0},\boldsymbol{z}]$,where $\boldsymbol{z} \in \mathcal{T}$ and $|\mathcal{T}|<+ \infty$ and the set $\mathcal{T}$ is the vertex set of the polyblock. 
        In addition, the vertex $\boldsymbol{z} \in \mathcal{T}$ is a proper vertex, if and only if it is not dominated by other vertices ($\boldsymbol{x}$ is dominated by $\boldsymbol{y}$ if $\boldsymbol{x} \preceq \boldsymbol{y}$). 
        The key idea of POA is to construct a series of polyblock to approximate the feasible set because the optimal solution of the problem in the polyblock is at one of the vertex of the polyblock. 

        \begin{definition}
            (Projection): Let $\mathcal{G} \subset \mathcal{R}_+^N$ be a non-empty normal set. 
            For any point $\boldsymbol{z} \in \mathcal{R}_+^N \setminus G$, connecting the line segment of $\boldsymbol{0}$ and $\boldsymbol{z}$ will intersect the of $\mathcal{G}$, i.e., $\partial^+ \mathcal{G}$, at the projection point $\pi_\mathcal{G}(\boldsymbol{z})$, where $\pi_\mathcal{G}(\boldsymbol  {z})=\lambda\boldsymbol{z},\lambda=\max\{\alpha>0|\alpha\boldsymbol{z} \in \mathcal{G}\}$.
        \end{definition}

        Based on the above definitions, for the polyblock $\mathcal{P}_k \supset \mathcal{G} \cap \mathcal{H}$, assuming that there is $\boldsymbol{v}_1 \in \mathcal{R}_{+}^{N} \setminus \mathcal{G}$, then $\mathcal{P}_k \setminus \mathcal{K}_{\pi_{\mathcal{G}}(\boldsymbol{v}_1)}^+$ still obtains $\mathcal{G} \cap \mathcal{H}$($\mathcal{K}_{\boldsymbol{x}}^+ =\{\boldsymbol{x}' \in \mathcal{R}_+^N | \boldsymbol{x}' \succ \boldsymbol{x}\}$), so the polyblock can be used to approximate the feasible set of the problem. 
        Specifically, the feasible region is approached step by step by generating a new polyblock from a simple polyblock, i.e., $\mathcal{P}_1 \supset \mathcal{P}_{2} \supset \cdots \supset \mathcal{G} \cap \mathcal{H}$.
        \begin{figure}[htbp]
            \renewcommand{\algorithmicrequire}{\textbf{Input:}}
            \renewcommand{\algorithmicensure}{\textbf{Output:}}
            \removelatexerror
            \begin{algorithm}[H]
                \label{algo:PA}
                \caption{The Optimal Power Allocation Algorithm}
                \begin{algorithmic}[1]
                    \REQUIRE $\{p_i^{max}\},{h_i},\sigma_\omega$,$a_i,b_i,c_i$,$\gamma_i^{min}$,$e_i^{min}$,$\delta$
                    \ENSURE $\boldsymbol{p}^*=\{p_i^*\},i \in \mathcal{M}$
                    \STATE \textbf{Initialization:} Construct an initial polyblock $\mathcal{P}_1$ containing the feasible domain $\mathcal{G} \cap \mathcal{H}$ with the vertex set $\mathcal{T}=\{\boldsymbol{z}\}$, $Q_0=-\infty$, $k=0$;
                    \REPEAT
                        \STATE $k=k+1$;
                        \STATE Select the optimal vertex from the vertex set $\mathcal{T}_k$: $z_k= \arg\max \{Q(\boldsymbol{z})|\boldsymbol{z} \in \mathcal{T}_k  \}$;
                        \STATE Calculate the projection of the vertex $\boldsymbol{z}_k$ on the boundary of $\mathcal{G}$: $\boldsymbol{x}_k=\pi_\mathcal{G}(\boldsymbol{z}_k)$;
                        \IF {$\boldsymbol{x}_k=\boldsymbol{z}_k$}
                            \STATE $\bar{\boldsymbol{x}}_k=\boldsymbol{z}_k$;
                        \ELSE
                            \IF {$\boldsymbol{x}_k \in \mathcal{G} \cap \mathcal{H}$ and $Q(\boldsymbol{x}_k) \geq Q_{k-1}$}
                                \STATE $\bar{\boldsymbol{x}}_k=\boldsymbol{x}_k$,$Q_k=Q(\boldsymbol{x}_k)$;
                            \ELSE
                                \STATE $\bar{\boldsymbol{x}}_k=\bar{\boldsymbol{x}}_{k-1}$, $Q_k=Q_{k-1}$;
                            \ENDIF
                            \STATE Construct the next vertex set:
                            \begin{equation*}
                                \begin{aligned}
                                    \mathcal{T}_{k+1}=&(\mathcal{T}_k \setminus \mathcal{T}_k^*)\\ &\cup \{\boldsymbol{z}_{k,i}=\boldsymbol{z}+(x_k^i-z^i)\boldsymbol{e}^i\\
                                    &|\boldsymbol{z} \in \mathcal{T}_k^*,i\in\{1,\cdots,N\}\}
                                \end{aligned}
                            \end{equation*}
                            \STATE Remove the improper vertices in $\mathcal{T}_{k+1}$ and the set $\{\boldsymbol{v} \in \mathcal{T}_{k+1}|\boldsymbol{v} \notin \mathcal{H}\}$.
                        \ENDIF
                    \UNTIL{$||\boldsymbol{z}_k-\bar{\boldsymbol{x}}_k|| \leq \delta$}
                    \STATE Let $\boldsymbol{x}^*=\bar{\boldsymbol{x}}_k$ to get the SINR solutions, and then use the relationship between SINR and power to obtain the corresponding power allocation $\boldsymbol{p}^*$.
                \end{algorithmic}
            \end{algorithm}
        \end{figure}

        In the $k$-th iteration, the optimal vertex $\boldsymbol{z}_k$ of the polyblock $\mathcal{P}_k$ is first calculated. 
        If $\boldsymbol{z}_k \in \mathcal{G}$, the algorithm ends. 
        Otherwise, its projection point $\pi_{\mathcal{G}}(\boldsymbol{z}_k)$ is calculated. 
        Since the set $\mathcal{G}$ is normal, the projection point can be solved by the binary search algorithm. 
        Suppose the projection point is $\boldsymbol{x}_k$, and use the projection point to generate a new polyblock $\mathcal{P}_{k+1}=\mathcal{P}_k \setminus \mathcal{K}_{\pi_{\mathcal{G}}(\boldsymbol{z}_k)}^+$ from the polyblock $\mathcal{P}_k$. The vertex set of the new polyblock $\mathcal{P}_{k+1}$ is $\mathcal{T}_{k+1}=(\mathcal{T}_k \setminus \mathcal{T}_k^*) \cup \{\boldsymbol{z}_{k,i}=\boldsymbol{z}+(x_k^i-z^i)\boldsymbol{e}^i|\boldsymbol{z} \in \mathcal{T}_k^*,i\in\{1,\cdots,N\}\}$, 
        where $\mathcal{T}_k^*=\{\boldsymbol{z} \in \mathcal{T}|\boldsymbol{z} \succ \boldsymbol{x}_k\}$. 
        Then by removing the improper vertices of $\mathcal{T}_{k+1}$, the vertex set of $\mathcal{P}_{k+1}$ can be obtained. 
        In practical applications, for the efficiency of the algorithm, the termination condition is usually set as $|\boldsymbol{z}_k-\boldsymbol{x}_k| \leq \delta$. 
        The final POA algorithm is summarized in the algorithm 1. 
        The $i$-th component of the initial polyblock's vertex is set as the max SINR that is available to the device $D_i$, i.e., $z^i=\frac{|h_i|^2 p_i^{max}}{\sigma_\omega^2}$. 
        Using the hidden monotonicity of the problem, our algorithm can obtain its optimal solution with lower complexity. 

    \section{Performance Evaluation}
        In this section, we present the simulation results of the proposed algorithm. 
        We consider a cell where devices are randomly distributed in a circle centered on the BS with a specified radius. 
        The channel gain between each device and the BS is assumed to be composed of two parts: the large-scale loss and Rayleigh-distributed small-scale fading. 
        The large-scale channel loss is $L_{i}=128.1+37.6\log_{10}(d_i[km])$ dB. 
        The Rayleigh fading coefficient follows an i.i.d. Gaussian distribution as $\beta_i \sim \mathcal{C}\mathcal{N}(0,1)$. 
        The noise power can be calculated by formula $\sigma_{\omega}^2=BN_0$, where the bandwidth is $B=180kHz$ and the noise power spectral density is $N_0=-174$ dBm/Hz. 
        We encode the standard video test sequences into SVC streams according to \cite{zhu2018scalable}. 
        
        Two reference schemes are compared with the proposed algorithm (NOMA-Proposed). 
        One is a NOMA-based scheme whose target is to maximize throughput of the system (NOMA-MT). 
        Here we optimize the power allocation to maximize the throughput under the same constraints as the proposed scheme. 
        The other is an OMA-based scheme (OMA), in which the wireless resource is allocated to maximize the QoS according to the gradient on the utility curve of the video, which is defined as the increase in utility with respect to the amount of resource required for that increase. 

        \begin{figure}[htbp]
            \subfigure[]{
                \centerline{\includegraphics[width=.6\columnwidth]{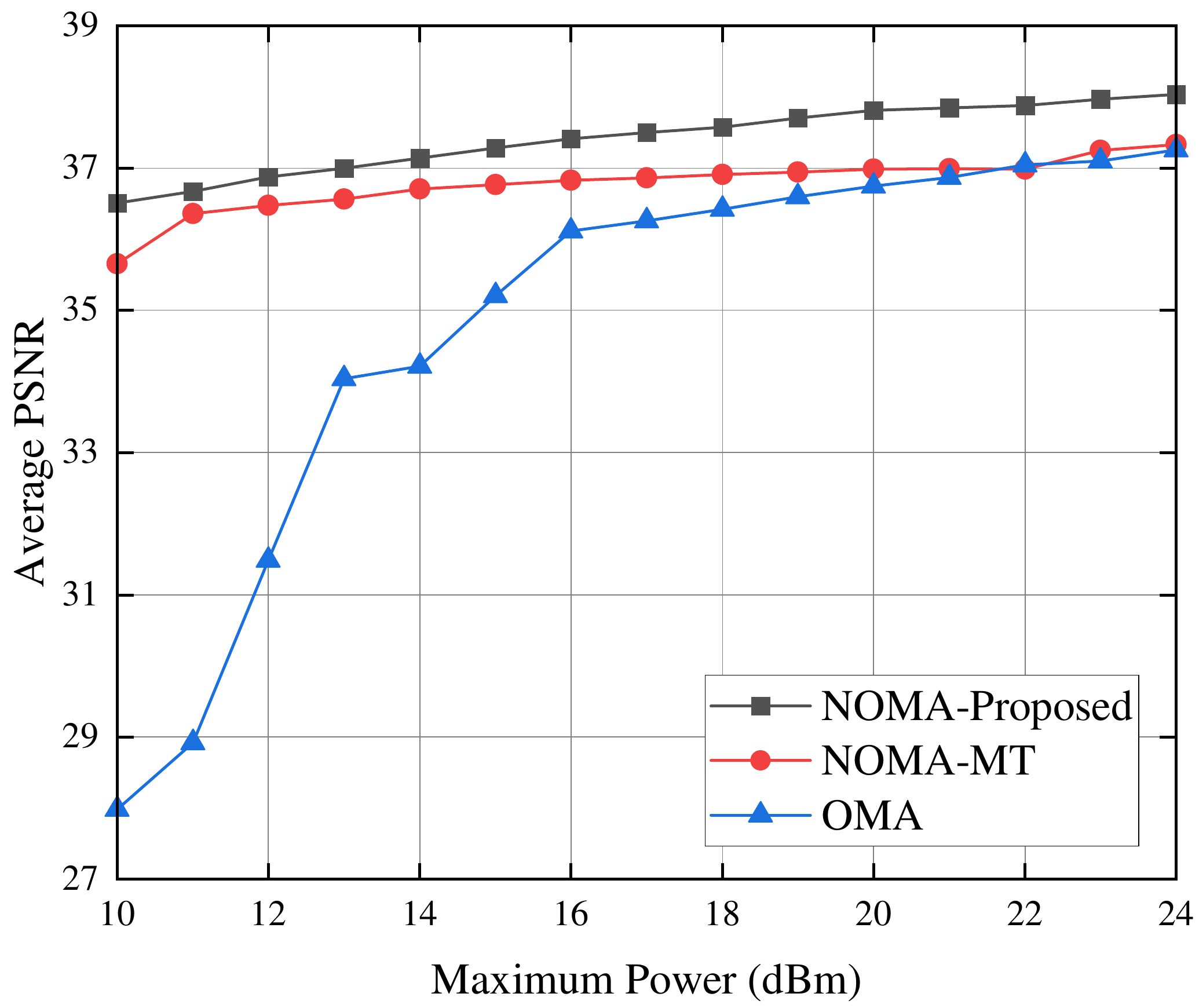}}
                \label{fig:p-psnr}
            }
            \subfigure[]{
                \centerline{\includegraphics[width=.6\columnwidth]{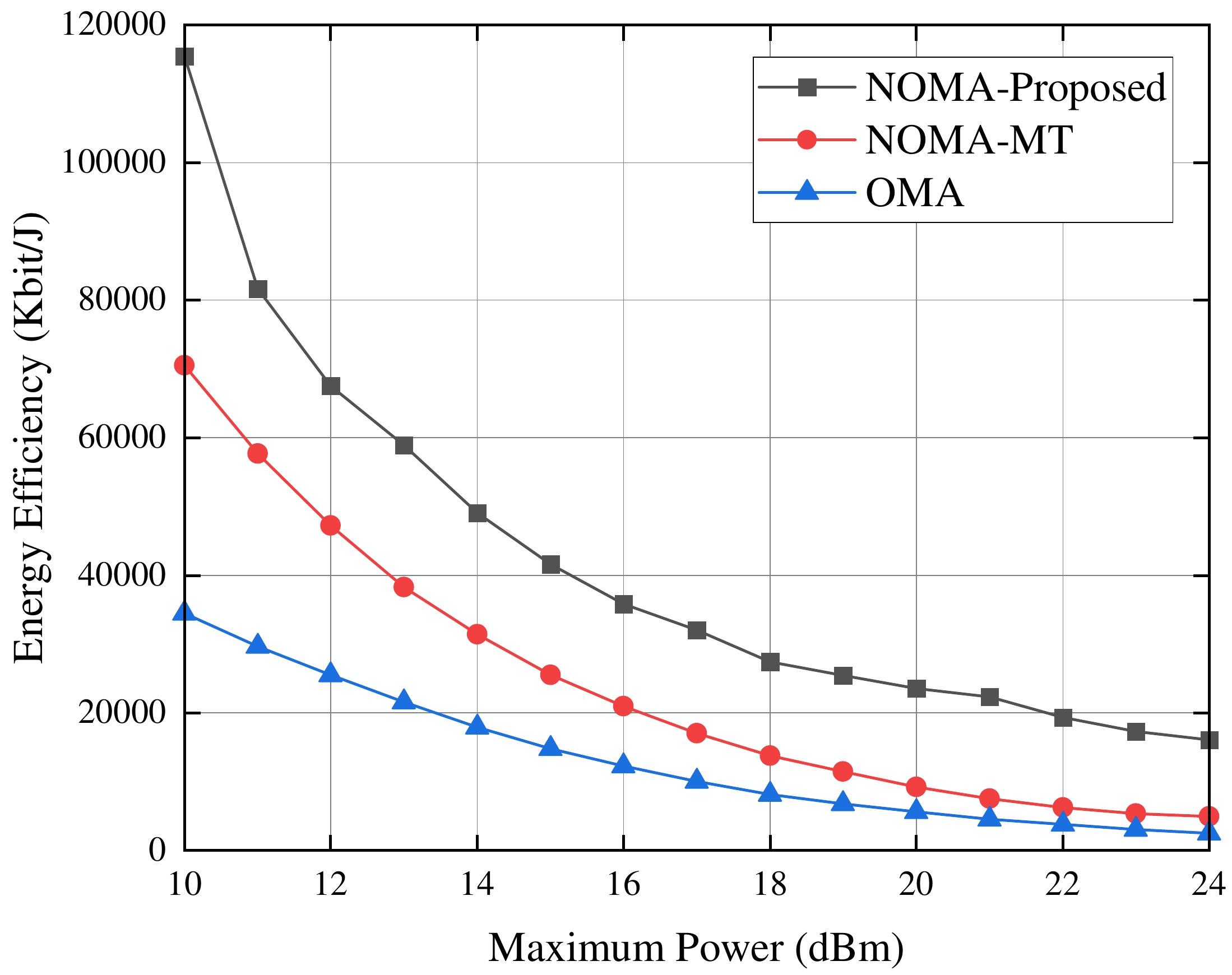}}
                \label{fig:p-ee}
            }
            \caption{(a) Average PSNR (QoS) vs. maximum transmission power with the service coverage = 1000 m; (b) Energy efficiency vs. maximum transmission power with the service coverage = 1000 m}
            \label{fig:power}
        \end{figure}
        First, we compare the QoS and EE of the three schemes, which is illustrated in Fig.~\ref{fig:power}. 
        We can observe that the NOMA-Proposed always performs better in both QoS and EE than others, as the maximum transmission power increases. 
        When the transmission power is relatively low, NOMA schemes can do much better than OMA scheme, because the high SE and high EE of NOMA allow schemes to transmit videos under limited bandwidth and power resource while OMA scheme fails to do so. 
        However, as the transmission power is high enough, traditional NOMA scheme that focuses on throughput can not maintain its advantage compare to OMA scheme while our proposed NOMA scheme still can show an improvement on QoS. 
        As for EE, it is obvious that NOMA schemes have a better performance than OMA scheme. 
        Besides, our proposed scheme also outperforms NOMA-MT due to our discrete QoS model can achieve a better utility of the transmission power. 

        \begin{figure}[htbp]
            \subfigure[]{
                \centerline{\includegraphics[width=.6\columnwidth]{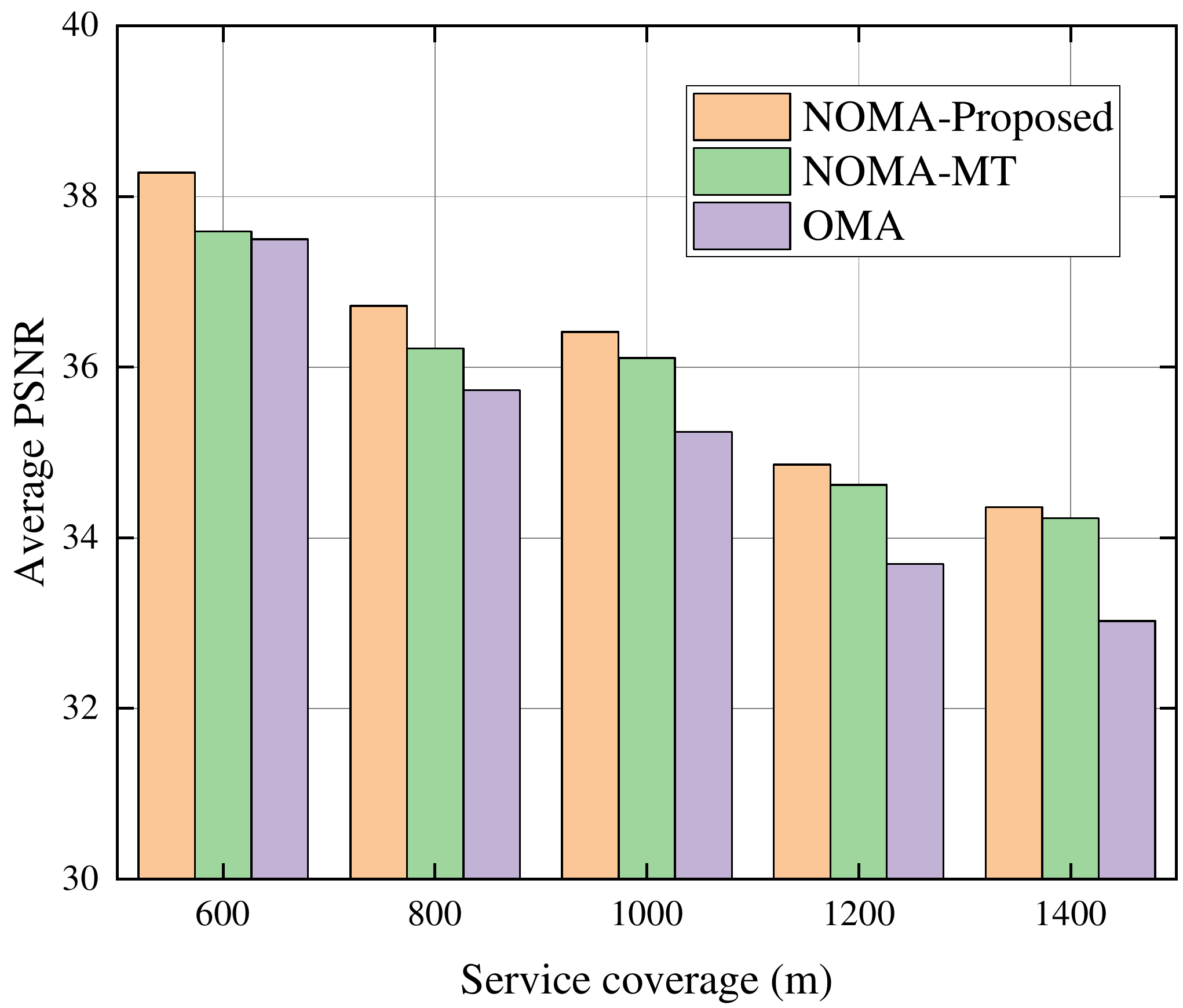}}
                \label{fig:d-psnr}
            }
            \subfigure[]{
                \centerline{\includegraphics[width=.6\columnwidth]{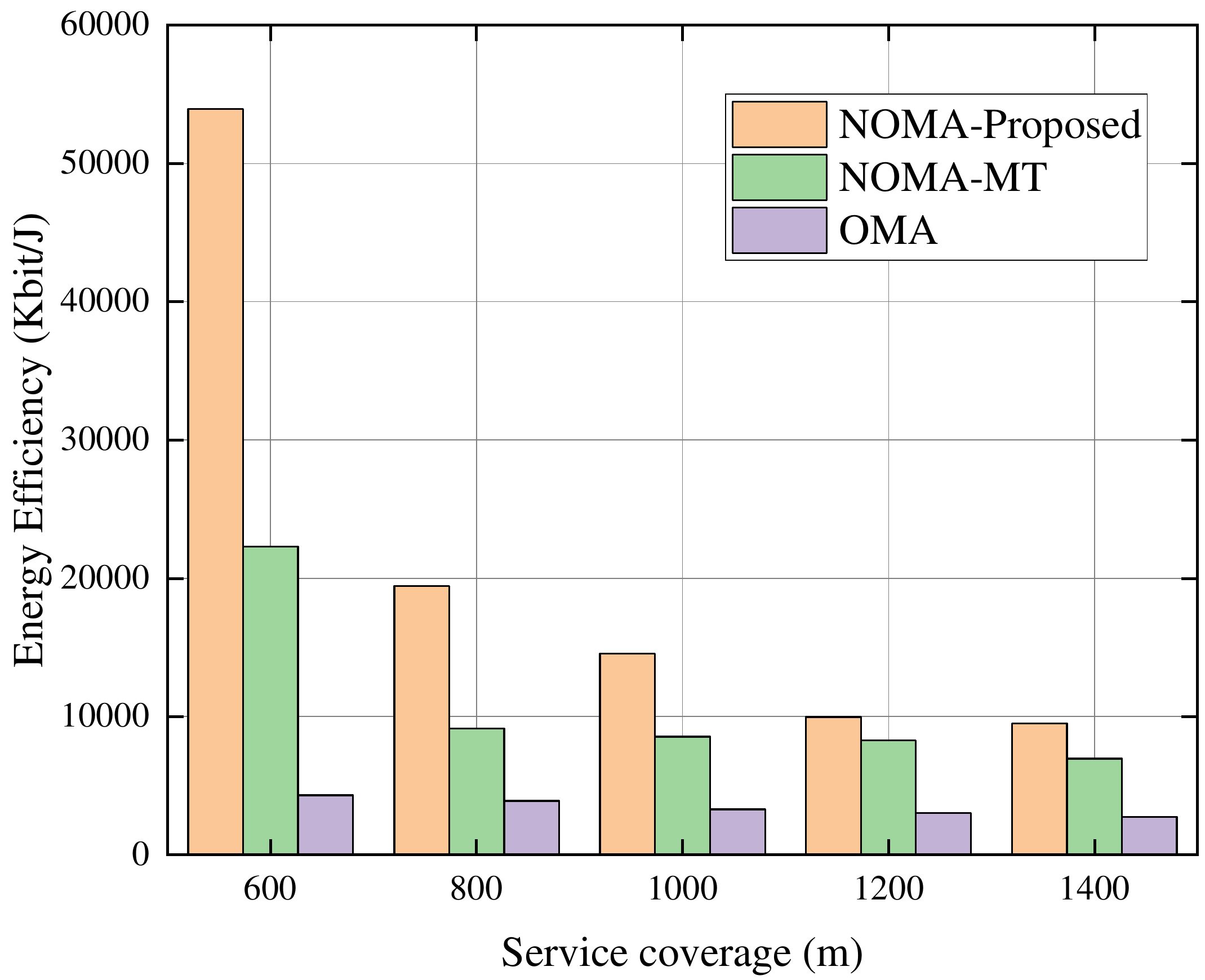}}
                \label{fig:d-ee}
            }
            \caption{(a) Average PSNR (QoS) vs. service coverage with the maximum power = 22dBm; (b) Energy efficiency vs. service coverage with the maximum power = 22dBm}
            \label{fig:distance}
        \end{figure}

        \begin{figure}[htbp]
            \centerline{\includegraphics[width=.6\columnwidth]{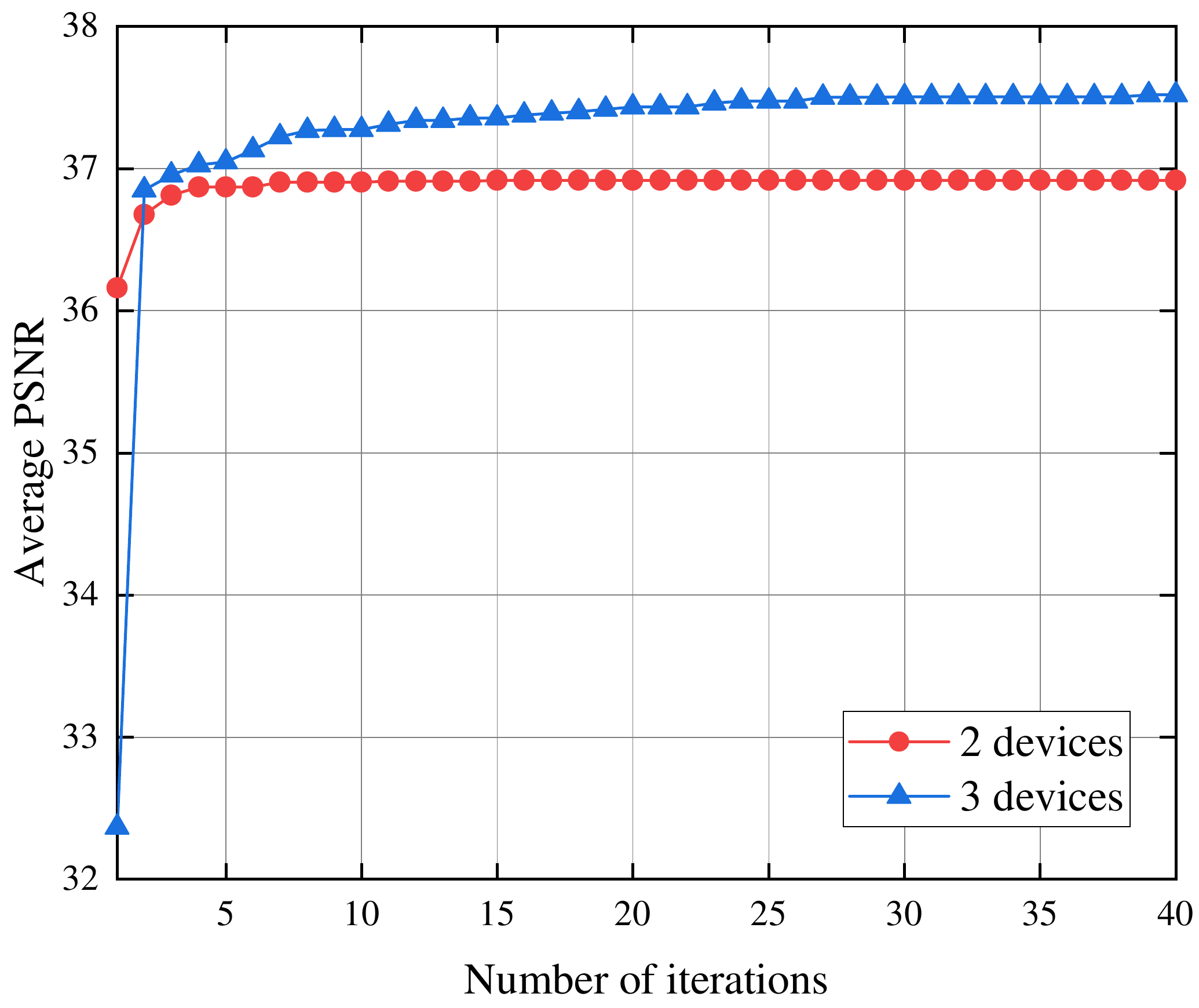}}
            \caption{Average PSNR vs. Number of iterations with the maximum power = 23dBm}
            \label{fig:iter-psnr}
        \end{figure}

        To demonstrate the anti damping performance of the NOMA-Proposed, we also evaluate the QoS and EE of these schemes as the service coverage increases from 600 to 1400 m. 
        The results in Fig.~\ref{fig:d-psnr} demonstrate that our proposed scheme still has a better QoS performance than the other schemes. 
        When the service coverage is small, the performance of traditional NOMA scheme is similar to that of OMA since the power is sufficient for both of them while our proposed scheme has a much better performance. 
        As the service coverage gets bigger, the advantage of NOMA becomes more significant. 
        And our proposed scheme also has a significant improvement on EE. 
        The gap of EE is not so big when the IoT devices are distributed in a large area. But when the devices get closer to the BS, the performance of our proposed scheme increases much more than NOMA-MT and OMA. This is because considerable energy is wasted as the achievable rate is far bigger than the video uplinking needs when the coverage is smaller and the channel state is better. 

        We also compare the coverage speed when the number of devices is different, which is shown in Fig.~\ref{fig:iter-psnr}. 
        We find out that when there are fewer devices, the algorithm converges faster since fewer devices means the solution space is much smaller.

    \section{Conclusion}
        In this paper, we propose an effective video uplinking scheme in NOMA-based IoT. 
        To make this scheme serve the video uplinking better, we establish a discrete QoS model. 
        Based on this QoS model, we formulate a QoS-driven optimization problem and design an optimal power allocation algorithm to solve the problem, whose performance is validated in simulations. 
        Considering the video uplinking in IoT, we make this first attempt to introduce the EE constraints into the problem. 
        Finally, simulation results show that the proposed scheme can provide a higher QoS and EE for video uplinking in IoT than other schemes.

    \section*{Acknowledgment}
        This work was supported in part by the National Science Foundation of China (NSFC) (Grants 61771445, 61631017 and U19B2044).

    \bibliographystyle{IEEEtran}
    \bibliography{paper-en}
\end{document}